\documentclass[12pt]{article}

\usepackage{graphicx}
\usepackage{amsmath}
\usepackage{amssymb}
\allowdisplaybreaks

\begin{document}

\baselineskip=17.5pt plus 0.2pt minus 0.1pt

\renewcommand{\theequation}{\thesection.\arabic{equation}}
\renewcommand{\thefootnote}{\fnsymbol{footnote}}
\makeatletter
\@addtoreset{equation}{section}
\def\CR{\nonumber \\}
\def\pt{\partial}
\def\be{\begin{equation}}
\def\ee{\end{equation}}
\def\bea{\begin{eqnarray}}
\def\eea{\end{eqnarray}}
\def\eq#1{(\ref{#1})}
\def\la{\langle}
\def\ra{\rangle}
\def\hyp{\hbox{-}}

\begin{titlepage}
\title{
\hfill\parbox{4cm}
{ \normalsize YITP-04-33 \\{\tt hep-th/0406167}}\\
\vspace{1cm}Evolving fuzzy $CP^n$ and lattice $n$-simplex}
\author{
Naoki {\sc Sasakura}\thanks{\tt sasakura@yukawa.kyoto-u.ac.jp}
\\[15pt]
{\it Yukawa Institute for Theoretical Physics, Kyoto University,}\\
{\it Kyoto 606-8502, Japan}}
\date{\normalsize June, 2004}
\maketitle
\thispagestyle{empty}

\begin{abstract}
\normalsize
Generalizing the previous works on evolving fuzzy two-sphere, I discuss evolving fuzzy $CP^n$
by studying scalar field theory on it. The space-time geometry is obtained in 
continuum limit, and is shown to saturate locally the cosmic holographic principle.
I also discuss evolving lattice $n$-simplex obtained by `compactifying' fuzzy $CP^n$.
It is argued that an evolving lattice $n$-simplex does not approach a continuum 
space-time but decompactifies into an evolving fuzzy $CP^n$.     
\end{abstract}
\end{titlepage}

\section{Introduction}
Several thought experiments in semi-classical quantum gravity and string theory show 
the existence of minimum length \cite{Garay:1994en,Yoneya:2000bt}. 
This suggests that space-time is not a continuous static object, 
but is fluctuating in essence and must be treated quantum mechanically in some way.
An approach to such quantum space-time is given by non-commutative geometry
\cite{Connes, Madore:aq}.
Although non-commutative geometry is technically much more complicated 
to treat than continuum geometry and its physical applications are quite limited at present, 
the approach is fascinating because space-time is naturally embedded 
in the operational properties of quantum field.

An interesting challenge to extend the applicability of non-commutative geometry would
be formulating dynamical evolutions of non-commutative spaces, as our universe is growing.
A general formulation of a fuzzy space changing its size \cite{Sasakura:2003ke} 
and field theory on evolving fuzzy two-sphere \cite{Sasakura:2004vm} were discussed
by the present author, based on pure-into-mixed state 
evolutions \cite{Hawking:1982dj, Banks:1983by}. Splitting process of fuzzy space was 
discussed in \cite{Balachandran:2003wv}. In fact these two processes are intimately related, 
and the evolution process can be interpreted as the process of a `main' fuzzy space 
emitting `baby' fuzzy spaces one after another \cite{Sasakura:2003ke, Sasakura:2004vm}.
Physically interesting is the space-time geometry associated with such an evolving fuzzy space.
The geometry on the evolving fuzzy two-sphere was extracted from the continuum limit of 
the scalar field theory on it \cite{Sasakura:2004vm}. It turned out that  
the behavior of the scale factor saturates locally the cosmic holographic principle proposed 
in \cite{Fischler:1998st}.  A certain class of observational 
limits on space-time quantities 
\cite{salecker,karolyhazy,Ng:1993jb,Amelino-Camelia:1994vs,Sasakura:1999xp} was argued 
\cite{Sasakura:1999xp,Ng:2004xr} to be related with the holographic principle 
\cite{'tHooft:gx, Susskind:1994vu,Fischler:1998st}. 
Therefore this property of saturation could specify the class
of space-time fuzziness associated with evolving fuzzy spaces.
 
The aim of the present paper is to extend part of the previous results on evolving fuzzy 
two-sphere to some other evolving fuzzy spaces. In Section \ref{cpn}, 
I will study scalar field theory on evolving fuzzy $CP^n$, and 
extract space-time geometry in continuum limit.
It will be observed that the space-time geometry saturates locally the cosmic holographic 
principle of \cite{Fischler:1998st}.
In Section \ref{simplex}, I will introduce potential terms into the action of scalar field
to `compactify' fuzzy $CP^n$ to lattice $n$-simplex, and discuss 
the space-time geometry in continuum limit. 
The final section will be devoted to summary and discussions.

\section{Evolving fuzzy $CP^n$}
\label{cpn}
For the present purpose of describing evolving fuzzy $CP^n$, it is most convenient
to introduce the Fock space generated by $n+1$ sets of creation-annihilation operators 
\cite{Alexanian:2001qj,Carow-Watamura:2004ct}. This description was used for evolving fuzzy 
two-sphere ($n$=1) in \cite{Sasakura:2004vm}. The creation-annihilation operators are defined by 
\be
\label{creationannihilation}
\begin{array}{l}
[a_i,a^\dagger_j]=\delta_{ij}, \cr
[a_i,a_j]=[a_i^\dagger,a_j^\dagger]=0,
\end{array} (i,j=1,\cdots,n+1).
\ee
The Fock space is constructed from the Fock vacuum $a_i \left| 0 \right> =0$ as
\be
\left|N;m_1,\cdots,m_{n+1}\right>=\frac{1}{\sqrt{m_1 !\cdots m_{n+1}!}}
(a_1^\dagger)^{m_1} \cdots (a_{n+1}^\dagger)^{m_{n+1}} \left| 0\right>,
\ee
where $m_i$ are natural numbers and a degenerate notation $N=\sum_{i=1}^{n+1} m_i$ is used 
for later convenience. The symmetry associated with $CP^n$ is $SU(n+1)$.
Its generators can be represented as operators on the Fock space,
\be
\label{defgen}
L^b=\sum_{i,j=1}^{n+1} a^\dagger_i T^b_{ij} a_j,
\ee
where $T^b_{ij}$ are the matrix elements of the hermitian generators in the fundamental 
representation of $SU(n+1)$. 
Let me define ${\cal H}_N$ as the subspace spanned by $\left|N;m_1,\cdots,m_{n+1}\right>$ with
any $m_i$ for fixed $N$. 
Each ${\cal H}_N$ gives the rank-$N$ symmetric representation
of $SU(n+1)$, and defines a fuzzy $CP^n$. The continuum limit is given by 
$N\rightarrow \infty$. A scalar field on a fuzzy $CP^n$ is an operator on ${\cal H}_N$, and
is denoted by $\phi_N$. I assume the hermiticity $\phi_N^\dagger=\phi_N$ for simplicity.

An evolution of a fuzzy space can be derived from unitary splitting process of
emitting `baby' fuzzy spaces \cite{Sasakura:2003ke, Sasakura:2004vm, Balachandran:2003wv}. 
For a fuzzy $CP^n$, its growing and shrinking processes are respectively described by 
the following $SU(n+1)$-invariant operations,
\bea
{\cal O}&\rightarrow& \sum_{i=1}^{n+1} a^\dagger_i {\cal O} a_i, \cr
{\cal O}&\rightarrow& \sum_{i=1}^{n+1} a_i {\cal O} a^\dagger_i,
\eea
where ${\cal O}$ is an operator on a fuzzy $CP^n$.
Generalizing the discussions about fuzzy two-sphere in \cite{Sasakura:2004vm} to the 
present case,
the equation of motion of a massless scalar field is given by 
\be
\label{eomscal}
\sum_{i=1}^{n+1} [a_i^\dagger,[a_i,\phi]]-\sum_{b} [L^b,[L^b,\phi]]=0,
\ee
where the summation of $b$ is over all the generators of $SU(n+1)$, and the scalar 
field is defined by $\phi=\sum_{N} \phi_N$.

To discuss the continuum limit of \eq{eomscal}, 
let me expand the scalar field in terms of the complete set of operators $Q_{j,m}^N$ as
\be
\label{expansionphi}
\phi=\sum_{N,j,m} \phi_{j,m}^N Q_{j,m}^N.
\ee
Here $j$ and $m$ label the second Casimir and an orthogonal coordinate in the subspace
labeled by $j$, respectively:
\bea
\label{defofq}
\sum_b [L^b,[L^b, Q^N_{j,m}]]&=&j \, Q_{j,m}^N, \cr
{\rm Tr}\left( {Q_{j,m}^N}^\dagger Q_{j',m'}^{N'}\right) &=& \delta_{NN'} \delta_{jj'}\delta_{mm'},
\eea
where the trace Tr is over the whole Fock space.
Because of the $SU(n+1)$ symmetry, the evolution operations can be assumed to have the form,
\bea
\label{qcoef}
\sum_{i=1}^{n+1} a^\dagger_i Q_{j,m}^N a_i &=& c_{j}^{N+} \, Q_{j,m}^{N+1}, \cr
\sum_{i=1}^{n+1} a_i Q_{j,m}^N a_i^\dagger &=& c_{j}^{N-} \, Q_{j,m}^{N-1},
\eea
with some coefficients $c_{j}^{N+}$ and $c_{j}^{N-}$. Note that the first equation of \eq{qcoef}
determines higher-rank orthogonal operators from lower ones. Therefore it is allowed to 
assume $c_j^{N+}$ be positive real.

To obtain $c_{j}^{N+}$, let me use the normalization condition in \eq{defofq},
\be
\label{cplus}
\left(c_j^{N+}\right)^2=\sum_{i,i'=1}^{n+1} {\rm Tr}\left(a_{i'}a_i^\dagger {Q_{j,m}^N} 
a_i a_{i'}^\dagger {Q_{j,m}^N}^\dagger\right).  
\ee
The $SU(n+1)$ generators in the fundamental representation satisfies the following 
identity 
\be
\label{tidentity}
\frac1{n+1} \delta_{\alpha\beta}\delta_{\gamma\delta}
+2 \sum_{b} T_{\alpha\beta}^bT_{\gamma\delta}^b=
\delta_{\alpha\delta}\delta_{\beta\gamma},
\ee 
where the standard normalization ${\rm tr}(T^a T^b)=\delta_{ab}/2$ is assumed.
Using this identity, the right-hand side of \eq{cplus} can be evaluated as 
\bea
(c_j^{N+})^2&=&\frac1{n+1} {\rm Tr}\left((N+n+1)Q_{j,m}^N(N+n+1){Q_{j,m}^N}^\dagger\right)
+2 \sum_{b} {\rm Tr}\left(L^b Q_{j,m}^N L^b {Q_{j,m}^N}^\dagger\right) \cr
&=& \frac{(N+n+1)^2}{n+1}+ 2\sum_b {\rm Tr}  \left( -\frac12[L^b,[L^b,Q_{j,m}^N]]{Q_{j,m}^N}^\dagger+
(L^b)^2  Q_{j,m}^N {Q_{j,m}^N}^\dagger \right) \cr
&=& (N+n+1)(N+1)-j.
\eea
Therefore
\be
c_j^{N+}=\sqrt{(N+n+1)(N+1)-j}.
\ee
The right-hand side of \eq{cplus} can be computed differently by using \eq{qcoef}. This 
gives
\be
\left(c_j^{N+}\right)^2=c_j^{N+} c_j^{(N+1)-}.
\ee
Therefore,
\be
c_{j}^{N-}=c_{j}^{(N-1)+}=\sqrt{N(N+n)-j}.
\ee

Substituting the expansion \eq{expansionphi} into the equation of motion \eq{eomscal},
I obtain 
\be
\sum_{N,j,m}
(-c_j^{(N-1)+}\phi_{j,m}^{N-1}-c_j^{(N+1)-}\phi_{j,m}^{N+1}+(2N+n+1-j)\phi_{j,m}^{N})Q_{j,m}^N=0.
\ee
I assume that this equation will approach 
a second order differential equation with respect to $N$ in the continuum limit. Then,
using the expansion,
\be
\phi_{j,m}^{N\pm 1}\simeq \phi_{j,m}(N) \pm \phi_{j,m}'(N) +\frac12 \phi_{j,m}''(N),
\ee
I obtain
\bea 
&&-\frac12(c_j^{(N-1)+}+c_j^{(N+1)-}){\phi_{j,m}''(N)}
+(c_j^{(N-1)+}-c_j^{(N+1)-})\phi_{j,m}'(N) \cr
&&\ \ \ \ \ \ \ \ \ \ \ \ \  +(2N+n+1-j-c_j^{(N-1)+}-c_j^{(N+1)-})\phi_{j,m}(N)=0,
\eea
where $N$ is regarded as a continuum variable and $'$ denotes the derivative with
respect to $N$.
Let me define 
\be
T=N+\frac{n+1}2,
\ee
and assume $N$ is large enough and
\be
\label{assjn}
\frac{j}{N}\ll 1.
\ee
Then
\bea
c_j^{(N-1)+}&\sim& T-\frac12, \cr
c_j^{(N+1)-}&\sim& T+\frac12,
\eea
and the equation of motion of a massless scalar field in the continuum limit 
is obtained as 
\be
\label{eomscalcon}
T \phi_{j,m}''(T)+\phi_{j,m}'(T)+j\ \phi_{j,m}(T)=0,
\ee
where $'$ denotes the derivative with respect to $T$. 
In a geometric background of
\be
\label{assummetric}
ds^2=-f(T) dT^2+ h(T) d\Omega^2,
\ee
where $d\Omega^2$ is the metric on a unit $CP^n$, the equation of motion of a massless 
scalar field takes the form,
\be
f(T)^{-1/2}h(T)^{-n+1} \frac{\pt}{\pt T} f(T)^{-1/2}h(T)^{n}  \frac{\pt}{\pt T} \phi(T,\Omega)
-\Delta_\Omega \phi(T,\Omega)=0, 
\ee
where $\Delta_\Omega$ denotes the Laplacian on a unit $CP^n$, and I have used the fact that
$CP^n$ has $2n$-dimensions. Comparing this with \eq{eomscalcon},
the eigenvalues of $\Delta_\Omega$ should be identified with $-j$, and
\bea
\label{idcon}
&&\frac{h(T)}{f(T)}=T, \cr
&&f(T)^{-1/2}h(T)^{-n+1} \frac{d}{d T} f(T)^{-1/2}h(T)^{n} =1.
\eea
The solution is unique up to an unimportant overall constant factor, and I obtain 
a metric, 
\bea
\label{metricscalcp}
ds^2&=&-T^{-\frac{n-1}{n-1/2}}dT^2 +T^\frac1{2n-1} d\Omega^2 \cr
&=&-T^{-\frac{D-2}{D-1}}dT^2+T^\frac1{D-1}d\Omega^2,
\eea
where $D=2n$, the dimensions of $CP^n$. 
Changing the time variable to $dt=T^{-\frac{D-2}{2D-2}}dT$, $t\sim T^\frac{D}{2D-2}$,
the metric of an evolving fuzzy
$CP^n$ in the continuum limit is given by
\be
\label{cpnmetric}
ds^2=-dt^2+t^{\frac2D}d\Omega^2
\ee
up to unimportant constants.  
 
For the consistency of the continuum limit, the cosmic time $t$ must be regarded
as a continuum variable. This is actually satisfied, since, for an interval of $\Delta N=1$,
\be
\Delta t\sim T^{-\frac{D-2}{2D-2}},
\ee
which remains constant for $D=2$ or vanishes for $D > 2$ in the limit $T\rightarrow \infty$. 
As for the spatial part, the number of `points' on a fuzzy $CP^n$ can be estimated 
by the dimensions of ${\cal H}_N$, which is approximately $N^{n}$.
Therefore, from the expression \eq{cpnmetric}, the volume per point is approximately 
given by 
\be
\sim \frac{(t^\frac1D)^D}{N^{n}}\sim t^{2-D},
\ee
which remains constant for $D=2$ or vanishes for $D > 2$ in the limit $T\rightarrow \infty$. 

From the equation of motion \eq{eomscalcon}, the angular frequency with 
respect to the variable $T$ of a solution is roughly $\omega_T\sim \sqrt{j/T} \sim \sqrt{j/N}$. 
Therefore
the assumption \eq{assjn} means $\omega_T \ll 1$. The physical angular 
frequency should be measured in terms of the variable $t$, and this is given by
$\omega_t =\omega_T \frac{dT}{dt}\sim \omega_T T^{\frac{D-2}{2D-2}}$. 
This $\omega_t$ can be identified with the physical energy of a mode and it has no upper bounds 
in the continuum limit for $D>2$, or must be much smaller than a constant for $D=2$. 
Hence the continuum limit is well-behaved also in this respect. 

In \cite{Fischler:1998st}, a cosmological version of the holographic principle 
\cite{'tHooft:gx,Susskind:1994vu} is proposed. This was obtained by imposing 
the holographic principle to the inner region causally connected with a boundary.
For a simple case of $a(t)\sim t^p$ for a spatially flat FRW universe $ds^2=-dt^2+a(t)^2 dx^2$, 
the cosmic holographic principle imposes an inequality
\be
\label{cosmichol}
p>\frac1D.
\ee
This is indeed saturated in \eq{cpnmetric}, if the curvature of $CP^n$ is ignored.
Therefore, concerning local properties, the present model saturates 
the cosmic holographic principle. This suggests that the present model is 
associated with the class of space-time fuzziness of 
\cite{salecker,karolyhazy,Ng:1993jb,Amelino-Camelia:1994vs,Sasakura:1999xp}, 
which was argued \cite{Sasakura:1999xp,Ng:2004xr}
to be related to the holographic principle 
\cite{'tHooft:gx, Susskind:1994vu}.   
On the other hand, as was discussed in \cite{Fischler:1998st} for a FRW closed universe, 
if the global structure of $CP^n$ is taken into account, the present model will be ruled out 
by the cosmic holographic principle. This is because a boundary can be quite small
with a large causally-connected inner region in $CP^n$.

\section{Evolving lattice $n$-simplex}
\label{simplex}

In this section, I will introduce potential terms to compactify fuzzy
$CP^n$ down to lattice $n$-simplex.
Similar methods were used in constructing fuzzy spaces from others in \cite{Nair:2003st,Abe}. 
The equation of motion of a scalar field in this section is given by
\be
\label{eomscallat}
k \sum_{i=1}^{n+1} [a_i^\dagger,[a_i,\phi]]-\sum_b [L^b,[L^b,\phi]]
-v \sum_{i=1}^{n+1} [N_i,[N_i,\phi]]=0,
\ee
where $k,v$ are positive real constants, and $N_i=a^\dagger_i a_i\ (i=1,\cdots,n+1)$. 
A nice property of these potential terms is that they commute with the kinetic terms,
\bea
[N_j,\sum_{i=1}^{n+1} [a_i^\dagger,[a_i,\phi]]] &=& \sum_{i=1}^{n+1} [a_i^\dagger,[a_i,[N_j,\phi]]], \cr
[N_j,\sum_b [L^b,[L^b,\phi]]]&=& \sum_b [L^b,[L^b,[N_j,\phi]]].
\eea
Therefore \eq{eomscallat} can be considered in each subspace of the eigenvalues of 
the adjoint operations $[N_j,\cdot]$.
Let me assume $v$ is so large that  
the physical degrees of freedom are effectively restricted to
\be
[N_{i},\phi]=0,\ (i=1,\cdots,n+1).
\ee
This allows an expansion,
\be
\label{phiexp}
\phi=\sum_{m_1,\cdots,m_{n+1}}\phi_{m_1,\cdots,m_{n+1}}\left|N;m_1,\cdots,m_{n+1}\right|,
\ee
where I have introduced a shorthand notation,
\be
\left|N;m_1,\cdots,m_{n+1}\right|\equiv\left|N;m_1,\cdots,m_{n+1}\right>
\left<N;m_1,\cdots,m_{n+1}\right|.
\ee

Plugging the expansion \eq{phiexp} into the first term of \eq{eomscallat}, I obtain
\bea
\label{firstterm}
\sum_{i=1}^{n+1}  [a_i^\dagger,[a_i,\phi]]&=&
-\sum_{i=1}^{n+1} (a_i^\dagger \phi a_i+a_i \phi a_i^\dagger)+(2N+n+1) \phi\cr
&=& \sum_{m_1,\cdots,m_{n+1}} 
\Bigg(-\sum_{i=1}^{n+1} \left(m_i \phi_{m_1,\cdots,m_i-1,\cdots,m_{n+1}}
+(m_i+1) \phi_{m_1,\cdots,m_i+1,\cdots,m_{n+1}}\right) \cr
&&\ \ \ \ \ \ \ +(2N+n+1) \phi_{m_1,\cdots,m_{n+1}}\Bigg) \left|N;m_1,\cdots,m_{n+1}\right|.
\eea
In the continuum limit where $m_i$ are large and can be regarded as continuum variables, 
I expand the scalar field in the second order of derivatives,
\be
\phi_{m_1,\cdots,m_i\pm 1,\cdots,m_{n+1}}\simeq\phi_c \pm \pt_i \phi_c+\frac12 \pt_i^2 \phi_c, 
\ee
where $\phi_c$ is the continuum limit of $\phi_{m_1,\cdots,m_{n+1}}$.
Substituting the expansion into \eq{firstterm} and assuming $m_i \gg 1$,
I obtain
\be
\label{estfirst}
\sum_{i=1}^{n+1} [a_i^\dagger,[a_i,\phi]] \simeq -\sum_{m_1,\cdots,m_{n+1}} \left(
\sum_{i=1}^{n+1} m_i \pt_i^2 \phi_c+ \pt_i \phi_c \right)
\left|N;m_1,\cdots,m_{n+1}\right|.
\ee

As for the second term in \eq{eomscallat}, 
from the expressions of the generators \eq{defgen}, the identity \eq{tidentity} and 
$[\sum_i N_i,\phi]=0$,
I obtain
\bea
\label{secndterm}
\sum_{b} [L^b,[L^b,\phi]]&=&
\frac 12 \sum_{i,j=1}^{n+1} [a_j^\dagger a_i,[a_i^\dagger a_j,\phi]] \cr
&=& \sum_{m_1,\cdots,m_{n+1}} \sum_{i,j=1}^{n+1} (m_i+1)m_j \left( \phi_{m_1,\cdots,m_{n+1}}
-\phi_{m_1,\cdots,m_i+1,\cdots,m_j-1,\cdots,m_{n+1}}\right) \cr
&& \ \ \ \ \ \ \ \ \ \ \ \ \ \ \ \ \times \left|N;m_1,\cdots,m_{n+1}\right|. 
\eea
In the continuum limit, I may use the expansion,
\be
\phi_{m_1,\cdots,m_i+1,\cdots,m_j-1,\cdots,m_{n+1}} \simeq
\phi_c+\pt_i \phi_c-\pt_j \phi_c +\frac12 \pt_i^2 \phi_c +\frac12 \pt_j^2 \phi_c-
\pt_i\pt_j\phi_c.
\ee
Then, substituting this expansion into \eq{secndterm} and assuming $m_i \gg 1$, I obtain
\bea
\label{estsecond}
\sum_{b} [L^b,[L^b,\phi]]&=&\sum_{m_1,\cdots,m_{n+1}}
\left(\sum_{i=1}^{n+1} ((n+1) m_i-N) \pt_i \phi_c+ \sum_{i=1}^{n+1} (m_i^2-m_i N)\pt_i^2\phi_c 
\right. \cr
&&\left. +\sum_{i,j=1,i\neq j}^{n+1} m_i m_j \pt_i\pt_j\phi_c\right) 
\left|N;m_1,\cdots,m_{n+1}\right|. 
\eea
Substituting \eq{estfirst} and \eq{estsecond} into \eq{eomscallat}, the equation of 
motion of a massless scalar field in the continuum limit is given by
\bea
\label{eomconlimitlat}
0&=&\sum_{i=1}^{n+1} (-k+N-(n+1) m_i) \pt_i \phi_c+\sum_{i=1}^{n+1} (-k m_i+ N m_i-m_i^2)
\pt_i^2\phi_c-\sum_{i,j=1,i\neq j}^{n+1} m_i m_j \pt_i\pt_j\phi_c \cr
&=& \sum_{i=1}^{n+1}\pt_i\left((-k m_i+N m_i-m_i^2)\pt_i\phi_c\right)
-\sum_{i,j=1,i\neq j}^{n+1} \pt_i\left(m_im_j\pt_j\phi_c\right).
\eea 
Here note that the orders of $m_i$ are different between 
the temporal (with $k$) and the spatial directions, i.e. one order is 
replaced with $k$.
The motivation for the introduction of the parameter $k$ in \eq{eomscallat} was
to avoid the confusions about the approximation in the continuum limit.
If the temporal terms were neglected as lower orders of $m_i$, the equation of motion would have
no propagation modes.

The equation of motion of a massless scalar field in a non-trivial geometric background
is given by
\be
\label{eqconscal}
\pt_i \sqrt{-g} g^{ij}\pt_j\phi=0.
\ee   
Identifying this with \eq{eomconlimitlat}, the geometry in the continuum limit is identified as
\be
\label{kmet}
\sqrt{-g}g^{ij}=\left\{ 
\begin{array}{ll}
-k m_i + N m_i-m_i^2 & {\rm for}\ i=j,\cr
-m_i m_j & {\rm for}\ i\neq j.
\end{array}
\right.
\ee
From \eq{kmet}, the determinant of the metric tensor is computed to be
\be
\label{detg}
-g=\left( k (N-k)^{n} \prod_{i=1}^{n+1} m_i\right)^{\frac2{n-1}}.
\ee
For $n=1$, \eq{kmet} leads to an inconsistency.
Therefore the geometric picture can be applied to the continuum limit only for $n>1$. 
This will be assumed in the discussions below.

To decouple the temporal and spatial directions, let me take a new parameterization,
\bea
m_i&=&N x_i, \ \ (i=1,\cdots,n) \cr
m_{n+1}&=&N \left(1-\sum_{i=1}^{n} x_i\right).
\eea
These coordinates satisfy $0<x_i,\sum_{i=1}^n x_i <1$. Denoting $N$ also by $x_0$ and assuming
$N,m_i\gg 1$, the metric tensor in this basis is given by
\bea
\label{latticemet}
\tilde g^{00}&=&- a N k, \cr
\tilde g^{0i}&=&0, \cr
\tilde g^{ij}&=& \left\{
\begin{array}{ll}
a x_i(1-x_i)& {\rm for}\ i=j \cr
-a x_i x_j & {\rm for}\ i\neq j 
\end{array}
\right.,
\eea
where 
\be
a=N^{-\frac{2n+1}{n-1}} \left(k\left(1-\sum_{i=1}^{n}x_i\right)\prod_{i=1}^{n}x_i
\right)^{-\frac1{n-1}}.
\ee

Extracting $N$ dependence from \eq{latticemet}, the metric has the form,
\be
\label{metlatext}
ds^2=-N^{\frac{n+2}{n-1}} A(x_i) dN^2 + N^{\frac{2n+1}{n-1}} d\Omega^2(x_i),
\ee
where $d\Omega^2(x_i)$ is the spatial part depending only on $x_i$.
Defining a cosmic time $dt=N^{\frac{n+2}{2(n-1)}}dN$, the metric can be expressed as 
\be
ds^2=-A(x_i) dt^2 + t^\frac{2(2n+1)}{3n} d\Omega^2(x_i).
\ee
Ignoring the spatial dependence, the exponent of the scale factor is consistent
with the cosmic holography \eq{cosmichol},
\be   
\frac{2n+1}{3n}-\frac1{n}=\frac{2(n-1)}{3n} >0.
\ee
 
The metric \eq{metlatext}, however, casts doubts on the consistency of the continuum limit. 
It is clear that the proper time of one unit of time
$\Delta N=1$ becomes infinitely larger, as $N$ becomes larger. As for the spatial
part, the volume per lattice is approximately given by 
$\big( N^{\frac{2n+1}{2(n-1)}}\big) ^{n}/N^{n}=N^\frac{3n}{2(n-1)}\sim t$, which
is also divergent.
To see more what is physically expected as $N$ becomes larger, 
let me go back to the starting equation of 
motion \eq{eomscallat}. 
If the excitation energy of the kinetic term is in the same
order of that of the potential term, the lattice compactification cannot be justified.
The lowest excitation of the potential term in the action \eq{eomscallat} is $v$,
while the order of the highest excitation of the kinetic term is given by $N^2$,
which is estimated by putting $\pt_i\sim O(1)$ and $m_i\sim O(N)$ in \eq{eomconlimitlat}.
Hence the decoupling cannot be justified for $N^2 > v$. Therefore, when $N$ becomes much
larger than $\sqrt{v}$, the compactification potential is  not large enough to keep the 
compactification, and the system is better described by a scalar field theory
on $CP^n$. 
 
\section{Summary and discussions}
In this paper, I have studied scalar field theory on an evolving fuzzy $CP^n$
and its compactified space, extending part of the previous results on an evolving fuzzy two-sphere 
\cite{Sasakura:2004vm}. The compactification to a lattice $n$-simplex was done
by introducing some potential terms. 
The space-time geometries in the continuum limit were extracted from scalar field theory, 
and it was observed that an evolving fuzzy $CP^n$ saturates locally 
the cosmic holographic principle \cite{Fischler:1998st}, while an evolving lattice $n$-simplex 
is consistent with it.
It was argued that an evolving lattice $n$-simplex does not approach a 
continuous space-time but decompactifies into an evolving $CP^n$. 

It is interesting that the holographic principle 
\cite{Fischler:1998st,'tHooft:gx, Susskind:1994vu} appears generally in the present models.
A class of observational limits on space-time quantities 
\cite{salecker,karolyhazy,Ng:1993jb,Amelino-Camelia:1994vs,Sasakura:1999xp} 
has been argued \cite{Sasakura:1999xp,Ng:2004xr} to be related to the holographic
principle.
These limits are based on the common property that longer observation leads 
to larger uncertainty. In fact, the present models should have a kind of information loss
through emitting `baby' fuzzy spaces. This loss should lead to larger uncertainty in longer
period of time. It would be worth studying the space-time fuzziness more directly by considering 
thought experiments in the present models.

Compactification of fuzzy spaces by potential terms or constraints \cite{Nair:2003st,Abe} 
is obviously an interesting new direction.
This will enlarge the varieties of fuzzy spaces. In the present models, however,
the compactified spaces do not evolve to continuum spaces but decompactifies into
the original ones. Though this process itself is also physically interesting, 
it could limit the varieties of continuum space-times 
which fuzzy spaces can evolve to.
Presently it is not clear whether there exist compactified fuzzy spaces
which approach continuum space-times.     

\vspace{.5cm}
\noindent
{\large\bf Acknowledgments}\\[.2cm]
The author was supported by the Grant-in-Aid for Scientific Research No.13135213 and No.16540244
from the Ministry of Education, Science, Sports and Culture of Japan.


\begin{thebibliography}{99}

\bibitem{Garay:1994en}
As a review, see L.~J.~Garay,
``Quantum gravity and minimum length,''
Int.\ J.\ Mod.\ Phys.\ A {\bf 10}, 145 (1995)
[arXiv:gr-qc/9403008].

\bibitem{Yoneya:2000bt}
For uncertainty relation in string theory, see
T.~Yoneya,
``String theory and space-time uncertainty principle,''
Prog.\ Theor.\ Phys.\  {\bf 103}, 1081 (2000)
[arXiv:hep-th/0004074].

\bibitem{Connes}
A.~Connes, ``Noncommutative Geometry,'' Academic Press (1994).

\bibitem{Madore:aq}
J.~Madore,
``An Introduction To Noncommutative Differential Geometry And Physical
Applications,''
Cambridge, UK: Univ. Pr. (2000) 371 p, (London Mathematical Society lecture note series. 257). 

\bibitem{Sasakura:2003ke}
N.~Sasakura,
``Non-unitary evolutions of noncommutative worlds with symmetry,''
JHEP {\bf 0401}, 016 (2004)
[arXiv:hep-th/0309035].

\bibitem{Sasakura:2004vm}
N.~Sasakura,
``Field theory on evolving fuzzy two-sphere,''
Class.\ Quant.\ Grav.\  {\bf 21}, 3593 (2004)
[arXiv:hep-th/0401079].

\bibitem{Hawking:1982dj}
S.~W.~Hawking,
``The Unpredictability Of Quantum Gravity,''
Commun.\ Math.\ Phys.\  {\bf 87}, 395 (1982).

\bibitem{Banks:1983by}
T.~Banks, L.~Susskind and M.~E.~Peskin,
``Difficulties For The Evolution Of Pure States Into Mixed States,''
Nucl.\ Phys.\ B {\bf 244}, 125 (1984).

\bibitem{Balachandran:2003wv}
A.~P.~Balachandran and S.~Kurkcuoglu,
``Topology change for fuzzy physics: Fuzzy spaces as Hopf algebras,''
arXiv:hep-th/0310026.

\bibitem{Fischler:1998st}
W.~Fischler and L.~Susskind,
``Holography and cosmology,''
arXiv:hep-th/9806039.

\bibitem{salecker}
H.\ Salecker and E.\ P.\ Wigner, ``Quantum Limitations of the Measurement of Space-Time Distances,''
 Phys.\ Rev.\ {\bf 109}, 571 (1958).

\bibitem{karolyhazy}
F.\ Karolyhazy, ``Gravitation and Quantum Mechanics of Macroscopic Objects",
Nuovo\ Cim.\ {\bf A42}, 390 (1966).

\bibitem{Ng:1993jb}
Y.~J.~Ng and H.~Van Dam,
``Limit to space-time measurement,''
Mod.\ Phys.\ Lett.\ A {\bf 9}, 335 (1994).

\bibitem{Amelino-Camelia:1994vs}
G.~Amelino-Camelia,
``Limits on the measurability of space-time distances in the semiclassical approximation of quantum gravity,''
Mod.\ Phys.\ Lett.\ A {\bf 9}, 3415 (1994)
[arXiv:gr-qc/9603014].

\bibitem{Sasakura:1999xp}
N.~Sasakura,
``An uncertainty relation of space-time,''
Prog.\ Theor.\ Phys.\  {\bf 102}, 169 (1999)
[arXiv:hep-th/9903146].

\bibitem{Ng:2004xr}
As a recent review, see Y.~J.~Ng,
``Quantum foam and quantum gravity phenomenology,''
arXiv:gr-qc/0405078.

\bibitem{'tHooft:gx}
G.~'t Hooft,
``Dimensional Reduction In Quantum Gravity,''
arXiv:gr-qc/9310026.

\bibitem{Susskind:1994vu}
L.~Susskind,
``The World as a hologram,''
J.\ Math.\ Phys.\  {\bf 36}, 6377 (1995)
[arXiv:hep-th/9409089].

\bibitem{Alexanian:2001qj}
G.~Alexanian, A.~P.~Balachandran, G.~Immirzi and B.~Ydri,
``Fuzzy CP(2),''
J.\ Geom.\ Phys.\  {\bf 42}, 28 (2002)
[arXiv:hep-th/0103023].

\bibitem{Carow-Watamura:2004ct}
U.~Carow-Watamura, H.~Steinacker and S.~Watamura,
``Monopole bundles over fuzzy complex projective spaces,''
arXiv:hep-th/0404130.

\bibitem{Nair:2003st}
V.~P.~Nair and S.~Randjbar-Daemi,
``Quantum Hall effect on S(3), edge states and fuzzy S(3)/Z(2),''
Nucl.\ Phys.\ B {\bf 679}, 447 (2004)
[arXiv:hep-th/0309212].

\bibitem{Abe}
Y.~Abe,
``A construction of fuzzy $S^4$,''
arXiv:hep-th/0406135.

\end{thebibliography}
\end{document}